\documentclass[figures]{article}
\usepackage{graphicx}
\usepackage{epsf}
\usepackage{bm}
\usepackage{amsmath}
\usepackage{amsthm}
\usepackage{amsfonts}
\usepackage{amssymb}
\usepackage{enumerate}

\begin{document}

\begin{center}
\LARGE\bf A New Theorem on the Nonclassicality of States
\end{center}

\begin{center}
\rm Farid Sh\"ahandeh\footnote{Corresponding author. E-mail: shahandeh@ikiu.ac.ir} \ and \ Mohammad Reza Bazrafkan
\end{center}

\begin{center}
\begin{footnotesize} \sl
\textit{Physics Department, Faculty of Science, I. K. I. University, Qazvin, Iran}
\end{footnotesize}
\end{center}

\vspace*{2mm}

\begin{abstract}
A new theorem on the non-classicality depth of states has been proved. We show that if $W_{\hat \rho} \left( \alpha_m,s_m \right)=0$ exist for some value of the ordering parameter $s$ at some phase-space point $\alpha_m$, and if $W_{\hat \rho} \left( \alpha,s_m \right)$ is an acceptable quasi-classical distribution, the non-classicality of $\hat \rho$ in parallel with Lee's non-classicality depth is then given by $\tau_m=\left( 1 - s_m \right)/2$. In this way, a general examination of the effects of the single-photon-addition and -subtraction operations has been studied. The theorem, indeed, provides a theoretical background for generating quantum states of arbitrary non-classicality depth.
\end{abstract}

\begin{center}
\begin{minipage}{15.5cm}
\begin{minipage}[t]{2.3cm}{\bf Keywords:}\end{minipage}
\begin{minipage}[t]{9.8cm}
Nonclassicality, General Ordering Theorem, $s$-Parameterized Operator Expansion Formula, Incomplete Two-Dimensional Hermite Polynomials.
\end{minipage}\par\vglue8pt
{\bf PACS: }
42.50.-p,~42.50.Dv,~03.65.Ta,~31.15.-p
\end{minipage}
\end{center}

\section{Introduction}

It is a long time physicists searching the similarities of quantum and classical mechanics. The task is interested due to several reasons such as its applications in testing foundations of quantum theory. However, now it became a last-longing question ``what do we mean exactly by a non-classical quantum state?'' Till now, a number of non-classicality measures of the states have been introduced. Perhaps Mandel's $Q$ parameter~\cite{Mandel} which measures the sub-Poissonian photon statistics, Hillary~\cite{Hillary1,Hillary2}, Dodonov \textit{et al}~\cite{Dodonov1,Dodonov2} and Marian \textit{et al}~\cite{Marian} distances which on different grounds measure how close is an state to some reference set of states, and Lee's $R$-function~\cite{Lee1} are the most distinguished criteria. The latter is closely related to the ordering problem of boson operators and the class of orderings introduced by Chahill and Glauber ~\cite{CG1,CG2}. Beginning from the coherent state representation~\cite{Glauber1,Sudarshan},
	\begin{equation}
	\hat \rho  = \int {\frac{{{d^2}\alpha }}{\pi }P\left( \alpha  \right)\left| \alpha  \right\rangle \left\langle \alpha  \right|},
	\end{equation}
Lee defines the function
	\begin{equation} \label{Rf}
	R\left( {\alpha ,\tau } \right) = \frac{1}{\tau }\int {\frac{{{d^2}\beta }}{\pi }{e^{ - \frac{{\left| {\beta  - \alpha } \right|}^2}{\tau }}}P\left( \beta  \right)},
	\end{equation}
in which the parameter $\tau$ represents the degree of the smoothness needed to convert the $P$-function into an acceptable quasi-classical phase-space distribution function, in the sense that the negativity or singularity of the $P$-function is considered to be a manifest of the non-classicality of the field state. The greatest lower bound, or infimum, of the parameter $\tau$ is then called the non-classicality depth of the state and has been denoted by $\tau_m$. Physically, this is related to the minimum photon number of the thermal noise needed to remove all the non-classical properties of the field. That is to say, addition of some thermal noise will turn the quasi-probability distribution of the state into a proper distribution within the framework of the classical stochastic theory.

To show the connection to the Cahill and Glauber $s$-parameter, observe that for two quasi-probability distributions one has the transformation~\cite{Glauber2}
	\begin{equation} \label{WsWt}
	{W_{\hat \rho }}\left( {\alpha ,s} \right) = \frac{2}{{t - s}}\int {\frac{{{d^2}\beta }}{\pi }{e^{ - \frac{{2{{\left| {\beta  - \alpha } \right|}^2}}}{{t - s}}}}{W_{\hat \rho }}\left( {\beta ,t} \right)},
	\end{equation}
which in the special case of $t=1$ gives
	\begin{equation} \label{Wst}
	{W_{\hat \rho }}\left( {\alpha ,s} \right) = \frac{2}{{1 - s}}\int {\frac{{{d^2}\beta }}{\pi }{e^{ - \frac{{2{{\left| {\beta  - \alpha } \right|}^2}}}{{1 - s}}}}P\left( \beta  \right)}.
	\end{equation}
A simple comparison of~\eqref{Wst} by~\eqref{Rf} gives $\tau  = \left( 1 - s \right) / 2$ or equivalently $ s = 1 - 2\tau$. This, indeed, leads to $R\left( {\alpha ,\tau } \right) = W\left( {\alpha ,1 - 2\tau } \right)$. For a discussion of the different domains of the parameter $s$ and their relation to the non-classicality depth of states see e.~g.~\cite{Lutkenhaus}. In this way, after solving the ordering problem for density operator, there remains a problem of finding its infimum which requires a consideration of the distribution function over the whole phase-space for all the values of the parameter $\tau$. Obviously, this is not generally a simple task to do depending on the nature of the operator $\hat \rho$. Nevertheless, the non-classicality depth of some basic quantum optical states have been evaluated. For example, Fock states with $n \geqslant 1$ have a depth of $\tau_m=1$~\cite{Lee1} to deserve to be realized as the most non-classical states. Malbouisson \textit{et al}~\cite{Malbouisson}  have also studied the superposition of states together with mixed states.

It is useful to remember an interesting theorem by Lee~\cite{Lee2} by which any state having no projection on the vacuum Fock state has the maximum non-classicality depth of $\tau_m=1$. From now on we use the equivalent values of the ordering parameter $s$, $s_m$, to refer to the non-classicality depth $\tau_m$.

In the present letter, we state and prove an extension of Lee's theorem. This provides a theoretical basis for constructing quantum states of \emph{arbitrary} non-classicality depth. In addition, in many cases, it will reduce the problem of finding infimum of the $s$-parameterized quasi-probability distribution over the whole complex plane to finding its zeros with respect to the parameter $s$ at a possible point $\alpha_m$. We also use the recently given theorem on the ordering of operators~\cite{SB} named as the general ordering theorem (GOT) together with the calculus of the incomplete 2-dimensional Hermite polynomials~\cite{Dattoli} to give an unadorned careful examination of the method in the general case of the single-photon-added and -subtracted states. We also discuss another remark on the depth of the non-classicality which we will call \textit{non-classicality degree}. This new notion will make a physical distinction between states of the same non-classicality depth for the known states at least.

\section{Ordered Fock Space Projectors and Lee's Truncated States}

According to Lee's theorem~\cite{Lee2} removing the vacuum part of any state $\hat \rho$ makes it to be as non-classical as possible. One might show this fact through the $t$-parameterized representation of the Fock states. This is given by~\cite{SBA}
	\begin{equation} \label{nm}
	\left| n \right\rangle \left\langle m \right| = \frac{1}{{\sqrt {n!m!} }}{f^{n + m + 1}}{\left\{ {{h_{n,m}}\left( {{a^\dag },a\left| \kappa  \right.} \right){e^{ - f{a^\dag }a}}} \right\}_t},
	\end{equation}
with
	\begin{equation}
	f = \frac{2}{{t + 1}},\qquad \kappa  = \frac{{{t^2} - 1}}{4}.
	\end{equation}
There exists a deep connection between the incomplete 2-D Hermite polynomials and the ordering problem for which Eq.~\eqref{nm} is just a special case. These functions are defined through the series~\cite{Dattoli}
	\begin{equation}
	{h_{n,m}}\left( {x,y|\kappa } \right) = \sum\limits_{j = 0}^{\min \left\{ {n,m} \right\}} {\left( {\begin{array}{*{20}{c}}
  n \\ 
  j 
\end{array}} \right)\left( {\begin{array}{*{20}{c}}
  m \\ 
  j 
\end{array}} \right)j!{\kappa ^j}{x^{n - j}}{y^{m - j}}},
	\end{equation}
and their generating function is given by
	\begin{equation}
	\sum\limits_{n,m = 0}^\infty  {\frac{{{\mu ^n}{\nu ^m}}}{{n!m!}}{h_{n,m}}\left( {x,y|\kappa } \right)}  = {e^{\mu x + \nu y + \kappa \mu \nu }}.
	\end{equation}
Using these functions one may write the ordering transformation formula~\cite{Wunsche}
	\begin{eqnarray}
	{\left\{ {{a^{\dag n}}{a^m}} \right\}_s} =& \sum\limits_{j = 0}^{\min \left\{ {n,m} \right\}} \left( {\begin{array}{*{20}{c}}
  n \\ 
  j 
\end{array}} \right)\left( {\begin{array}{*{20}{c}}
  m \\ 
  j 
\end{array}} \right)j! \nonumber\\
& \times {{\left( {\frac{{t - s}}{2}} \right)}^j}{{\left\{ {{a^{\dag n - j}}{a^{m - j}}} \right\}}_t}
	\end{eqnarray}
in the simple form
	\begin{eqnarray}
	{\left\{ {{a^{\dag n}}{a^m}} \right\}_s} = {\left\{ {{h_{n,m}}\left( {{a^\dag },a|{\kappa _{s,t}}} \right)} \right\}_t}, \label{st} \\ {\kappa _{s,t}} = \frac{{t - s}}{2}.
	\end{eqnarray}

In the special case of the vacuum projector, using Eq.~\eqref{nm}, its $t$-ordered form is then given by
	\begin{equation}
	\left| 0 \right\rangle \left\langle 0 \right| = \frac{2}{{t + 1}}{\left\{ {{e^{ - \frac{{2{a^\dag }a}}{{t + 1}}}}} \right\}_t},
	\end{equation}
which leads to the $\left(-t\right)$-parameterized quasi-probability distribution of
	\begin{equation}
	{W_{\left| 0 \right\rangle \left\langle 0 \right|}}\left( {\alpha , - t} \right) = \frac{2}{{t + 1}}{e^{ - \frac{{2{{\left| \alpha  \right|}^2}}}{{t + 1}}}}.
	\end{equation}
This obviously represents the positivity of the vacuum state's phase-space distribution for the values of the parameter $t \geqslant -1$. Now, one may simply use the Fock state representation of the density operator $\hat \rho  = \sum\limits_{n,m} {{\rho _{nm}}\left| n \right\rangle \left\langle m \right|} $ to write its vacuum part as
	\begin{equation} \label{rho00}
	{\rho _{00}} = {\text{Tr}}\left( {\hat \rho \left| 0 \right\rangle \left\langle 0 \right|} \right) = \frac{2}{{t+ 1}}\int {\frac{{{d^2}\alpha }}{\pi }{e^{ - \frac{{2{{\left| \alpha  \right|}^2}}}{{t + 1}}}}W_{\hat \rho}\left( {\alpha ,t} \right)},
	\end{equation}
where we have used the well-known trace formula in the phase-space representation~\cite{Glauber2},
	\begin{equation} \label{trace}
	{\text{Tr}}\left( {\hat F\hat G} \right) = \int {\frac{{{d^2}\alpha }}{\pi }{W_{\hat F}}\left( {\alpha ,s} \right){W_{\hat G}}\left( {\alpha , - s} \right)}.
	\end{equation}

\begin{figure}
	\begin{center}
	\includegraphics[width=6cm,height=1cm]{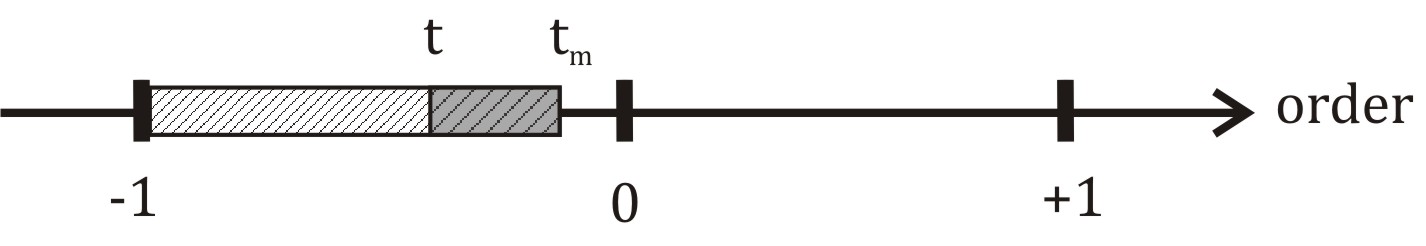}
	\end{center}
	\caption{$t_m>-1$ and $W_{\hat \rho}\left( {\alpha ,t} \right)$ being an acceptable classical distribution function imply $\rho_{00}> 0$ for all $t \geqslant -1$.}
	\label{f.W1t}
	\end{figure}

One may simply check that putting $t=1-2\tau$ into Eq.~\eqref{rho00} recovers Eq.~(12) of Ref.~\cite{Lee2}. If there exist some value $t_m>-1$ for which the quasi-probability distribution function $W_{\hat \rho}\left( {\alpha ,t} \right)$  is an acceptable quasi-classical one, then this will be true for all the values $-1 \leqslant t \leqslant t_m$ (see, e. g. Fig.~\ref{f.W1t}).
Thus, Eq.~\eqref{rho00} which is a Gaussian convolution formula implies that once $t_m>-1$ and $W_{\hat \rho}\left( {\alpha ,t} \right)$ is an acceptable quasi-classical distribution then $\rho_{00}> 0$ for all $t \geqslant -1$. The only way to make this therm vanish and $W_{\hat \rho}\left( {\alpha ,t} \right)$ to be still an acceptable classical distribution function is to have $t_m=-1$ and $W_{\hat \rho}\left( {0 ,-1} \right)=0$. This, indeed, represents the most non-classical state. It is shown by Jones \textit{et al}~\cite{Jones} that one can simply produce this \textit{truncated states} by adding just one photon to the field state.

\section{The Zero Point} In this section we state and proof a theorem extending the idea of Lee.

\newtheorem*{ZPT}{Theorem}
\begin{ZPT}
Consider the $s$-parameterized quasi-probability distribution function corresponding to some quantum state $\hat \rho$ is $W_{\hat \rho}\left( {\alpha ,s} \right)$. If
	\begin{enumerate}
	\item for some value of the ordering parameter, $s_m$, one finds $W_{\hat \rho}\left( {0 ,s_m} \right)=0$,
	\item and $W_{\hat \rho}\left( {\alpha ,s_m} \right)$ is an acceptable quasi-classical distribution,
	\end{enumerate}
then the non-classicality depth of $\hat \rho$ is $\tau_m=\left( 1 - s_m \right)/2$. Moreover, if
	\begin{enumerate}
	\item for some value of the ordering parameter, $s_m$, and at some phase-space point, $\beta_m$, one finds $W_{\hat \rho}\left( {\beta_m ,s_m} \right)=0$,
	\item and $W_{\hat \rho}\left( {\alpha ,s_m} \right)$ is an acceptable quasi-classical distribution,
	\end{enumerate}
then the non-classicality depth of $\hat \rho$ is $\tau_m=\left( 1 - s_m \right)/2$.
\end{ZPT}

\newtheorem*{ZPTP}{Proof}
\begin{ZPTP}
We may treat the $\left(-s\right)$-ordered form of the density operator in a similar way to that of Fock state representation. In this regard, we use the expansion $\hat \rho={\hat \rho ^{\left( { - s} \right)}} = \sum\limits_{n,m} {\rho _{nm}^{\left( { - s} \right)}{{\left\{ {{a^{\dag n}}{a^m}} \right\}}_{ - s}}} $ together with Eq.~\eqref{st} to write
	\begin{equation} \label{rhost}
	{\hat \rho ^{\left( { - s} \right)}} = \sum\limits_{n,m} {\rho _{nm}^{\left( { - s} \right)}{{\left\{ {{h_{n,m}}\left( {{a^\dag },a|{\kappa _{ - s, - t}}} \right)} \right\}}_{ - t}}},
	\end{equation}
which immediately gives the $t$-parameterized quasi-probability distribution of $\hat \rho$ as
	\begin{equation}
	W_{\hat \rho}\left( {\alpha ,t} \right) = \sum\limits_{n,m} {\rho _{nm}^{\left( { - s} \right)}{h_{n,m}}\left( {{\alpha ^ * },\alpha |{\kappa _{ - s, - t}}} \right)}.
	\end{equation}
We are interested in the evaluation of the coefficients $\rho _{nm}^{\left( { - s} \right)}$. To this end, we employ Eqs.~\eqref{nm} and~\eqref{trace} and write
	\begin{eqnarray}
	{\text{Tr}}\left( {\left| n \right\rangle \left\langle {m}
 \mathrel{\left | {\vphantom {m k}}
 \right. \kern-\nulldelimiterspace}
 {k} \right\rangle \left\langle l \right|} \right) &=& \frac{1}{{n!m!}}{\left( { - \frac{1}{\kappa }} \right)^{n + m + 1}} \nonumber \\
&&\qquad \times \int {\frac{{{d^2}\alpha }}{\pi }{e^{\frac{{{{\left| \alpha  \right|}^2}}}{\kappa }}}{h_{n,m}}\left( {{\alpha ^ * },\alpha \left| \kappa  \right.} \right){h_{k,l}}\left( {{\alpha ^ * },\alpha \left| \kappa  \right.} \right)} \nonumber \\ 
 &=& {\delta _{m,k}}{\delta _{n,l}}.
	\end{eqnarray}
Then this simply gives the expansion coefficients as
	\begin{equation}
	\rho _{nm}^{\left( { - s} \right)} = \frac{1}{{n!m!}}{\left( { \frac{1}{\kappa_{-t,-s} }} \right)^{n + m + 1}}\int {\frac{{{d^2}\alpha }}{\pi }{e^{ - \frac{{{\left| \alpha  \right|}^2}}{\kappa_{-t,-s}}}}{{h_{n,m}}\left( {{\alpha ^ * },\alpha \left| \kappa_{-s,-t}  \right.} \right)}W_{\hat \rho}\left( {\alpha ,t} \right)}.
	\end{equation}
For the present purpose, the first coefficient, namely $\rho^{\left(-s\right)}_{00}$, is sufficient. So we have
	\begin{equation} \label{rho00s}
	\rho _{00}^{\left( { - s} \right)} = \frac{2}{{t - s}}\int {\frac{{{d^2}\alpha }}{\pi }{e^{ - \frac{{2{{\left| \alpha  \right|}^2}}}{{t - s}}}}W_{\hat \rho}\left( {\alpha ,t} \right)}.
	\end{equation}
Interestingly, in the case of $s=-1$, or just the $Q$-function, one arrives at
	\begin{equation}
	\rho _{00}^{\left( 1 \right)} = \frac{2}{{t + 1}}\int {\frac{{{d^2}\alpha }}{\pi }{e^{ - \frac{{2{{\left| \alpha  \right|}^2}}}{{t + 1}}}}W_{\hat \rho}\left( {\alpha ,t} \right)},
	\end{equation}
which is the same as Eq.~\eqref{rho00}. In other words, the vacuum projection of the field state is just the zero order term in the expansion of the $Q$-function.

If there exist some infimum value $t_m>-1$ for which the quasi-probability distribution function $W_{\hat \rho}\left( {\alpha ,t_m} \right)$  is an acceptable quasi-classical one, then this will be true for all the values $-1 \leqslant t \leqslant t_m$. Equation~\eqref{rho00s} implies that once $t_m>-1$ and $W_{\hat \rho}\left( {\alpha ,t} \right)$ is an acceptable quasi-classical distribution then $\rho_{00}^{\left( -s \right)}> 0$ for all $t \geqslant -1$ and $-1 \leqslant s \leqslant t$ (see, e. g. Fig.~\ref{f.Wst}).
	\begin{figure}
	\begin{center}
	\includegraphics[width=6cm,height=1cm]{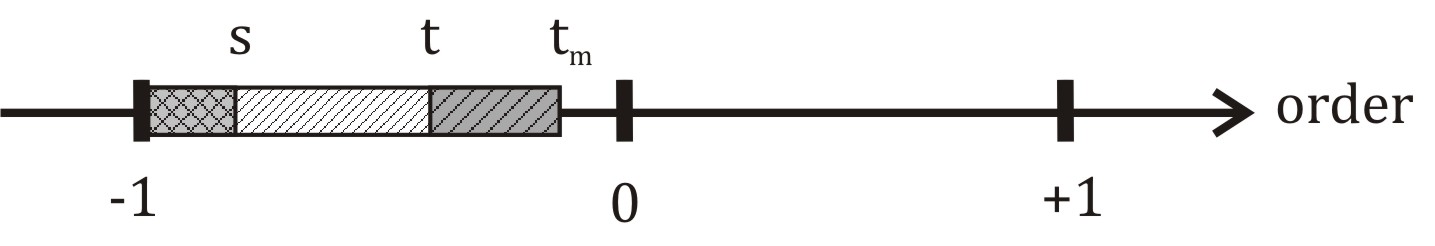}
	\end{center}
	\caption{$t_m>-1$ and $W_{\hat \rho}\left( {\alpha ,t} \right)$ being an acceptable classical distribution function imply $\rho_{00}^{\left( -s \right)}> 0$ for all $t \geqslant -1$ and $-1 \leqslant s \leqslant t$.}
	\label{f.Wst}
	\end{figure}
The only way to make this therm vanish and $W_{\hat \rho}\left( {\alpha ,t} \right)$ to be still an acceptable quasi-classical distribution function is to have $t_m=s$ and $W_{\hat \rho}\left( {0 ,s} \right)=0$. This, indeed, concludes the proof of the first part of the theorem.
\sloppy

The proof of the second part is straightforward.  This might be done by shifting the center of the convolution kernel in Eq.~\eqref{rho00s} to some other point $\beta$. Then, we have
	\begin{equation} \label{rhop00s}
	\rho_{00}^{\left( { - s} \right)} {\left( \beta \right)} = \frac{2}{{t - s}}\int {\frac{{{d^2}\alpha }}{\pi }{e^{ - \frac{{2{{\left| \alpha - \beta  \right|}^2}}}{{t - s}}}}W_{\hat \rho}\left( {\alpha ,t} \right)},
	\end{equation}
which is just the zero order term of the $\left( -s \right)$-ordered expansion of the displaced state $\hat \rho \left( \beta  \right) \equiv {{\hat D}^\dag }\left( \beta  \right)\hat \rho \hat D\left( \beta  \right)$. Now, in one hand we may use the previous reasoning to conclude that $\rho_{00}^{\left( { - s_m} \right)} {\left( \beta \right)} = 0$ and $W_{\hat \rho \left( \beta \right)}\left( {\alpha ,s_m} \right)$ being an acceptable quasi-classical distribution function imply $\tau_m = \left( 1 - s_m \right)/2$. On the other hand, a comparison of the Eq.~\eqref{rhop00s} with Eq.~\eqref{WsWt} leads to $\rho_{00}^{\left( { - s_m} \right)} {\left( \beta \right)} = W_{\hat \rho}\left( {\beta ,s_m} \right)$ and thus the result.
\end{ZPTP}
\fussy

\section{Applications}

The first example of the theorem given in the previous section is to consider the Fock states. Equation~\eqref{nm} leads to the $s$-parameterized phase-space distribution of the Fock states at zero to be
	\begin{equation} \label{Wn0}
	{W_{\left| n \right\rangle \left\langle n \right|}}\left( {0,s} \right) = {\left( { - 1} \right)^{2n + 1}}{\left( {\frac{{s + 1}}{2}} \right)^n}{\left( {\frac{2}{{s - 1}}} \right)^{n + 1}},
	\end{equation}
with the obvious root of $s_m=-1$ for all $n>0$ which implies that Fock states have the maximum non-classicality depth. This is so because $s=-1$ corresponds to the $Q$-function which is always positive allover the phase-space plane. The only possible quasi-classical Fock state is $n=0$ or just the vacuum state for which there exist no finite roots. One may examine the coherent states as well. The $s$-parameterized quasi-probability distribution function of the coherent projectors has been given by Fan~\cite{Fan} to be
	\begin{equation}
	{W_{\left| \alpha  \right\rangle \left\langle \alpha  \right|}}\left( {\beta ,s} \right) = \frac{2}{{1 - s}}{e^{ - \frac{{2{{\left| \alpha - \beta  \right|}^2}}}{{1 - s}}}}.
	\end{equation}
In this case, $W_{\left| \alpha  \right\rangle \left\langle \alpha  \right|}\left( {0 ,s} \right)$ possesses no zeros, and thus we cannot conclude anything about the nonclassicality depth of a coherent state using the zero point theorem above.

For the next example, we will consider the general effect of a single photon addition. To this end, employing the GOT~\cite{SB} we can simply write the following lemma;
	\begin{eqnarray}
	{a^k}{{\hat \rho }^{\left( { - s} \right)}} &=& {a^k}\sum\limits_{n,m} {\rho _{nm}^{\left( { - s} \right)}{{\left\{ {{a^{\dag n}}{a^m}} \right\}}_{ - s}}}  \nonumber\\
&=& \sum\limits_{n,m} \rho _{nm}^{\left( { - s} \right)}\sum\limits_{j = 0}^{\min \left\{ {n,m} \right\}} \left( {\begin{array}{*{20}{c}}
  k \\ 
  j 
\end{array}} \right)\left( {\begin{array}{*{20}{c}}
  n \\ 
  j 
\end{array}} \right)j! \nonumber \\
&\times& {{\left( {\frac{{1 - s}}{2}} \right)}^j}{{\left\{ {{a^{\dag n - j}}{a^{m + k - j}}} \right\}}_{ - s}}  \nonumber \\
&=& {\left\{ {{{\left[ {a + \left( {\frac{{1 - s}}{2}} \right){\partial _{{a^\dag }}}} \right]}^k}{{\hat \rho }^{\left( { - s} \right)}}} \right\}_{ - s}}.
	\label{L1}
	\end{eqnarray}
In a similar way, one obtains
	\begin{eqnarray}
	{{\hat \rho }^{\left( { - s} \right)}}{a^k} = {\left\{ {{{\left[ {a - \left( {\frac{{1 + s}}{2}} \right){\partial _{{a^\dag }}}} \right]}^k}{{\hat \rho }^{\left( { - s} \right)}}} \right\}_{ - s}}, \label{L2}\\
{a^{\dag k}}{{\hat \rho }^{\left( { - s} \right)}} = {\left\{ {{{\left[ {{a^\dag } - \left( {\frac{{1 + s}}{2}} \right){\partial _a}} \right]}^k}{{\hat \rho }^{\left( { - s} \right)}}} \right\}_{ - s}}, \label{L3}\\
{{\hat \rho }^{\left( { - s} \right)}}{a^{\dag k}} = {\left\{ {{{\left[ {{a^\dag } + \left( {\frac{{1 - s}}{2}} \right){\partial _a}} \right]}^k}{{\hat \rho }^{\left( { - s} \right)}}} \right\}_{ - s}}. \label{L4}
	\end{eqnarray}
Applying these for a single photon addition we obtain
	\begin{eqnarray}
	{a^\dag }{{\hat \rho }^{\left( { - s} \right)}}a = \sum\limits_{n,m} \rho _{nm}^{\left( { - s} \right)}\left\{ {a^{\dag n + 1}}{a^{m + 1}} + \left( {m + n + 1} \right){\kappa _{1,-s}}{a^{\dag n}}{a^m} \right. \nonumber \\ \left. + mn\kappa _{1,-s}^2{a^{\dag n - 1}}{a^{m - 1}} \right\}_{ - s}.
	\end{eqnarray}
This gives the zero order term as ${\kappa _{1,-s}}\rho _{00}^{\left( { - s} \right)} + \kappa _{1,-s}^2\rho _{11}^{\left( { - s} \right)}$ and thus $W_{a^{\dag}\hat \rho a}\left( {0 ,s} \right)$ will have a zero at $s_m=-1$. Then, provided that $W \left( {0 ,s} \right)$ has no singularity at $s=-1$, adding a single photon to the field will completely turn it into the most non-classical state. This is of course what we expected from Lee's theorem as it had been considered by Jones \textit{et al}~\cite{Jones}. An example of this case is photon-added coherent state which has a non-classicality depth of $s_m=-1$.

Similarly, an examination of the single-photon-subtracted states leads to the conclusion that the zero order term is ${\kappa _{-1,-s}}\rho _{00}^{\left( { - s} \right)} + \kappa _{-1,-s}^2\rho _{11}^{\left( { - s} \right)}$. This gives $s_m=1$, if the state does not posses a singularity at $\alpha=0$ and $s=1$ and that $\tau_m=0$ when $W_{\hat \rho} \left( \alpha,1 \right)$ is an acceptable quasi-classical distribution function as pointed out by Kim \textit{et al}~\cite{Kim}. For instance, in the case of Fock states or coherent states, they all have singularities at $\alpha = 0$ and $s = 1$ so that we can not decide on their non-classicality after the action of a single photon subtraction. However, we may get surprised considering the single-photon-subtracted thermal state (SPSTS). In this case, the state does not posses any singularities for all the values $ - \infty  \leqslant s \leqslant 1 + 2\left\langle {{n_{th}}} \right\rangle $. One may use the $P$-function of a thermal state to write its anti-normally ordered form as $ \vdots \exp \left( { - {a^\dag }a/\left\langle {{n_{th}}} \right\rangle } \right)/\pi \left\langle {{n_{th}}} \right\rangle  \vdots $. Then, using the general transformation rule ${\left\{ {\exp \left( {\lambda {a^\dag }a} \right)} \right\}_s} = 2{\left\{ {\exp \left[ {2\lambda {a^\dag }a/\left( {2 - \lambda t + \lambda s} \right)} \right]} \right\}_t}/\left( {2 - \lambda t + \lambda s} \right)$~\cite{SB} and lemmas \eqref{L1} and~\eqref{L4}, she can simply write the unnormalized $s$-parameterized quasi-probability function of SPSTS as
	\begin{eqnarray}
	{W_{a{{\hat \rho }_{th}}{a^\dag }}}\left( {\alpha ,s} \right) &=& \frac{2}{{\pi \left( {2\left\langle {{n_{th}}} \right\rangle  + 1 - s} \right)}}\nonumber \\ &&\times \left\{ {\left[ {\frac{{2\left\langle {{n_{th}}} \right\rangle  - 1 + s}}{{2\left\langle {{n_{th}}} \right\rangle  + 1 - s}} + {{\left( {\frac{{1 - s}}{{2\left\langle {{n_{th}}} \right\rangle  + 1 - s}}} \right)}^2}} \right]{{\left| \alpha  \right|}^2}} \right. \nonumber\\
	&&\left. { + \left( {\frac{{1 - s}}{2}} \right)\left( {\frac{{2\left\langle {{n_{th}}} \right\rangle }}{{2\left\langle {{n_{th}}} \right\rangle  + 1 - s}}} \right)} \right\}{e^{ - \frac{{2{{\left| \alpha  \right|}^2}}}{{2\left\langle {{n_{th}}} \right\rangle  + 1 - s}}}}.
	\end{eqnarray}
After considering its value just at $\alpha=0$ one readily concludes that its non-classicality depth is $s_m=1$. That is to say, a thermal state which has a well-behaved $P$-function will encounter a decrease of classical properties after just a single photon subtraction. This is, of course, what was predicted by the zero point theorem above.

It is also remarkable that the action of photon subtracting-then-adding will generally produce states of non-classicality depth of $s_m=-1$~\cite{SYLee}.

\section{Conclusion and Discussion}

According to what just stated, one may introduce \textit{the degree of non-classicality}, $\iota$, to be the degree of the zero of ${W_{\hat \rho }}\left( {0,s} \right)$ at $s_m$. For example, in view of Eq.~\eqref{Wn0}, each Fock state $\left| n \right\rangle$ has a non-classicality degree of $\iota = n$. Physically, This might be interpreted as the number of photon subtractions needed to make a Fock state quasi-classical. In this way, even though all Fock states with $n>0$ having the same non-classicality depth, they are different in the degree of non-classicality. The same is true for $n$ photon-added states: They all will have a non-classicality depth of $s=-1$, however, their degrees of non-classicality are different so that $\iota = n$. The effect of this degree is not apparent to us, or we do not have a general idea about the procedures reducing or increasing $\iota$. The only obvious cases are that of Fock states for which we know photon addition and subtraction will increase and decrease this parameter, and that of photon-added states for which photon addition will increase their degree of non-classicality. Nevertheless, we hope the definition come into practice.

In summary, we have shown that the vacuum projection of any state is equivalent to the value of the $Q$-function at $\alpha = 0$. In addition, we have proved a theorem showing that one may simply evaluate the non-classicality depth of a given state through the zeros of its $s$-parameterized phase-space distribution with respect to the ordering parameter $s$, provided that it exists. That is to say, if there exist the values $\alpha_m$ and $s_m$ satisfying ${W_{\hat \rho }}\left( {\alpha_m ,s_m} \right)=0$ and ${W_{\hat \rho }}\left( {\alpha ,s_m} \right)$ being an acceptable probability distribution, then this uniquely determines the non-classicality depth of the state to be $\tau_m=\left( 1 - s_m \right)/2$. Using GOT we have explicitly shown that adding just one photon to any field with a well-behaved $Q$-function at $\alpha=0$ completely ruins its quasi-classical structure. An interesting case of the single-photon-subtraction was also studied for which we have seen a decrease in the classical properties of the thermal state. In addition, we have given the distinction between seemingly equivalent fully non-classical states introducing the concept of degree of non-classicality.

To conclude, we may note that the theorem given here provides a theoretical basis for quantum state engineering of \emph{arbitrary} non-classicality depth. In other words, the theorem implicitly states that if, in some way, one could remove the value of the $s$-parameterized quasi-probability distribution of some quasi-classical state at some phase-space point, then she could obtain a state with non-classicality depth of $s$. However, there still remains the problem of whether there exists a practical procedure to obtain such states or not.

\end{document}